\documentclass[a4paper,onecolumn,11pt,unpublished]{quantumarticle}
\pdfoutput=1
\usepackage[utf8]{inputenc}
\usepackage[english]{babel}
\usepackage[T1]{fontenc}

\usepackage[usenames,dvipsnames]{color}
\usepackage[hyperfootnotes=false]{hyperref}
\hypersetup{
	colorlinks,
	citecolor=quantumviolet,
	linkcolor=quantumviolet,
	urlcolor=quantumviolet}
\usepackage{tikz}
\usepackage{lipsum}
\usepackage[numbers,sort&compress]{natbib}

\usepackage{algorithmicx}
\usepackage[linesnumbered,ruled,vlined]{algorithm2e}
\usepackage[noend]{algpseudocode}
\usepackage{mathtools}
\usepackage{graphicx}
\usepackage{amssymb}
\usepackage{amsmath}
\usepackage{amsthm}
\usepackage{amsfonts}
\usepackage{color}
\usepackage{xcolor}
\usepackage{braket}
\usepackage{soul}

\usepackage{subfigure}
\usepackage{booktabs}
\usepackage{hyperref}

\SetKwInput{KwInput}{Input}                
\SetKwInput{KwOutput}{Output}              

\DeclareMathOperator{\tr}{\mathrm{tr}} 
\DeclareMathOperator*{\argmax}{\mathrm{arg\,max}}

\definecolor{CCTLABgreen}{RGB}{0,128,0}

\begin{document}

\title{Self-Guided Quantum State Learning for Mixed States }

\author{Ahmad Farooq}
\orcid{0000-0002-6090-9553}
\author{Muhammad Asad Ullah}
\orcid{0000-0002-2494-3914}
\author{Syahri Ramadhani}
\author{Junaid ur Rehman}
\orcid{0000-0002-2933-8609}
\author{Hyundong Shin}
\orcid{0000-0003-3364-8084}
\affiliation{Department of Electronics and Information Convergence Engineering, Kyung Hee University, Korea}
\email{hshin@khu.ac.kr}
\thanks{(corresponding author)}
\maketitle

\begin{abstract}
We provide an adaptive learning algorithm for tomography of general quantum states. Our proposal is based on the simultaneous perturbation stochastic approximation algorithm and is applicable on mixed qudit states. The salient features of our algorithm are efficient ($O \left( d^3 \right)$) post-processing in the dimension $d$ of the state, robustness against measurement and channel noise, and improved infidelity performance as compared to the contemporary adaptive state learning algorithms. A higher resilience against measurement noise makes our algorithm suitable for noisy intermediate-scale quantum applications.
\end{abstract}

\section{Introduction}

We require high-fidelity state preparation and its characterization for applications in quantum communication \cite{GT:07:NP,BDCZ:98:PRL}, quantum computation \cite{DiV:95:Sci,NC:02:AA}, and quantum metrology \cite{GLM:11:NP,DRC:17:RMP}. 
Quantum state tomography is the task of inferring a quantum state from the statistics of measurements on multiple copies of the quantum system---which requires large computation and storage space.
The tomography algorithm resorts often to least-square estimation \cite{OWV:97:PRA}, maximum likelihood estimation \cite{BDPS:99:PRA,TZEFH:11:PRL}, hedged maximum likelihood estimation \cite{Koh:10:PRL}, Bayesian mean estimation \cite{Rob:10:NJP,GCC:16:NJP,HH:12:PRA,KSRHHK:13:PRA}, and linear regression estimation \cite{QQHLDXG:13:SR,QHWDZLXWLG:17:njpQI}.

Some recent proposals for quantum state tomography include self-guided quantum tomography (SGQT), practical adaptive quantum tomography (PAQT), and state tomography through eigenstate extraction with neural networks \cite{Fer:14:PRL, CFP:16:PRL, GFF:17:NJP, MGN:20:PRA,RQKFWR:21:PRL,ATND:19:SR}. 
SGQT employs a stochastic approximation optimization technique known as simultaneous perturbation stochastic approximation (SPSA) \cite{Spa:92:ITAC} to learn an unknown \emph{pure} state. This online machine learning technique converges to the desired unknown state extremely fast and requires much fewer amount of space and time as compared to other known standard quantum tomography algorithms. Furthermore, it is robust to noise, which is a highly desirable property in near-term quantum devices. However, its restriction to pure states seriously limits its utility in general scenarios where we often encounter mixed quantum states. The PAQT algorithm aims to remove this limitation by generalizing it to the mixed states. It achieves this feat by applying Bayesian mean estimation on data generated by the SGQT \cite{GFF:17:NJP}. The PAQT inherits the aforementioned desirable properties of the SGQT including efficiency in storage space and computation and robustness to noise.

In this paper, we propose a quantum state tomography algorithm, called the self-guided quantum state learning for mixed state, to estimate mixed quantum states of dimension $d$. The main ingredient of our algorithm is the observation that the SPSA algorithm converges to the eigenvector corresponding to the highest eigenvalue of a given unknown mixed state. We perform the complete characterization of an unknown mixed state by iteratively invoking the SPSA on the intersection of nullspace of already obtained eigenvectors. After estimating all eigenvectors and eigenvalues, we employ the Gram-Schmidt process to generate an orthonormal spectrum of the unknown state. This procedure produces accurate estimates of the unknown state with efficient post-processing and superior robustness to noise as illustrated by the numerical examples. Importantly, if the unknown given quantum state is pure, our algorithm is equivalent to the SGQT algorithm. Thus, inheriting all the desirable properties of SGQT for the pure state tomography.

The remainder of this paper is organized as follows. In Section~2, we outline our algorithm. Numerical simulation results are provided in Section~3. This is followed by experimental results on IBM quantum (IBMQ) devices in Section~4. Conclusion and some possible future directions are provided in Section~5. 

\section{Method}
A quantum state of $d$ dimension is defined by the density matrix $\rho=\sum_{i=1}^d p_{i}\ket{\psi_{i}}\bra{\psi_{i}}$ where $p_i$'s are the ordered probabilities (eigenvalues) such that  $0\leq  p_d  \leq 
\ldots  \leq p_2 \leq p_{1} \leq 1$ and  $\sum_{i=1}^d p_{i}=1$. Let $r$ be the rank of the density matrix $\rho$. Then, $p_{r+1} = p_{r+2} = \ldots = p_d =0$. 
To extract the information from a quantum system, we need to perform the measurements. The outcomes of the measurement follow the probability distribution $\{ P_i\}_i$, which is characterized 
through the Born's rule \cite{NC:02:AA} 
\begin{equation} \label{eq:1}
	P_i = \tr\left(\ket{\phi_{i}}\bra{\phi_{i}} \rho\right)=\bra{\phi_{i}}\rho\ket{\phi_{i}}.
\end{equation}
The elementary measurement operators  are the collection of positive operators, which sums to the identity operator $\sum_{i}\ket{\phi_{i}}\bra{\phi_{i}}=I$ in the $d$-dimensional Hilbert space. 

\begin{figure}[t!]
	\centering
	\includegraphics[width = 1 \textwidth]{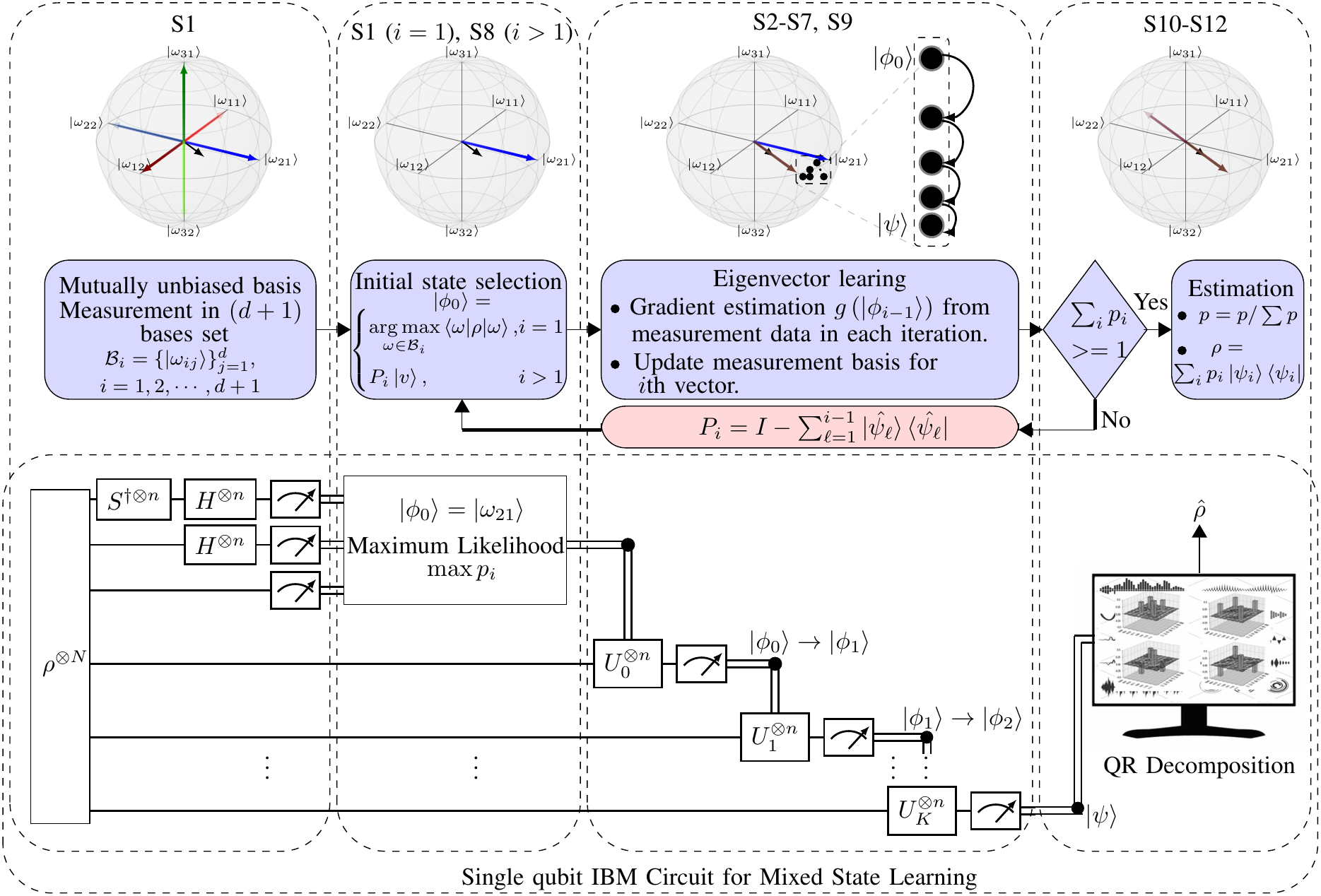}
	\caption{Self-guided quantum state learning for mixed states: The upper panel shows the general working of our algorithm. The state vectors (blue, green, and red) of MU basis and the unknown state (black) are shown in the Bloch sphere in S1. For estimating the first eigenvector ($i = 1$), the initial state vector $\ket{\phi_{0}}$ is selected from the basis state that has the maximum expectation on the unknown quantum state. This initial choice is followed by $K$ iterations of stochastic gradient estimation and state update. For estimating the remaining eigenvectors ($i > 1$), the initial choice is a random vector in the intersection of nullspace of previously estimated eigenvectors. Finally, the obtained probabilities are normalized and the estimate $\hat{\rho}$ is produced. See Section~2 for details. The lower panel shows the quantum circuit for implementation of our algorithm. See Section~4 for details.   
	} \label{fig:1}
\end{figure}

The SPSA is a stochastic algorithm to obtain the minimum solution of differential equations. In our case, the roots are the eigenvectors $\ket{\psi_{i}}$ of the density matrix $\rho$. To sequentially estimate the eigenvectors (eigenstates) of $\rho$, we design the SPSA to tackle the following optimization problem:
\begin{align}
	\begin{split}
	&	
	\argmax_{\ket{\phi}}  \quad \quad \bra{\phi}\rho\ket{\phi} \\
&	\text{subject to}\quad~  \mathrm{tr}\left(\ket{\phi}\bra{\phi}\right)=1 \\ & \hspace{2.1cm} \ket{\phi}\bra{\phi}\geq 0,
	\end{split}
\end{align}
which enables to estimate the most dominant eigenvector $\ket{\psi_{1}}$ of $\rho$ corresponding to the largest eigenvalue $p_1$.
We provide a sketch of our algorithm to estimate the density matrix $\rho$ of a mixed quantum state (see Figure~\ref{fig:1}). 

\begin{enumerate}
	
	\item[(S1)]
	Initialize the learning state vector  $\ket{\phi_{0}}$ to estimate the first eigenvector $\ket{\psi_{1}}$: Let $\left(d+1\right)$ 
	bases $\left\{\ket{\omega_{ij}}\right\}_{j=1}^d$, $i=1,2,\ldots,d+1$, be mutually unbiased (MU), enabling to optimally determine the density matrix $\rho$ of an ensemble of $d$ dimensional systems with informational completeness \cite{KF:89:AP,RFS:20:JSAC}.	Since these MU bases have the equal inner-product property across all mutual sets such that $\left| \braket{\omega_{ij} | \omega_{kl}} \right|^2=1/d$ for all $i \neq k, j, l$,\footnote{
	See \cite{KF:89:AP} for a construction of complete sets of MU bases.}  
we divide the (hyper)space of quantum states into $d\left(d+1\right)$ regions of basis states $\ket{\omega_{ij}}$ in the MU bases to initialize the learning state vector  $\ket{\phi_{0}}$.
	We first form the initializing set $\mathcal{B}_1 = \left\{ \ket{\omega_{ij}} \right\}$, then measure the density matrix $\rho$ in all the complete $d\left(d+1\right)$ MU projectors $\ket{\omega_{ij}}\bra{\omega_{ij}}$, and choose the basis state with the maximum probability as the starting state vector $\ket{\phi_{0}}$ to learn  the first eigenvector $\ket{\psi_{1}}$ as follows:
\begin{align}
	\ket{\phi_{0}}
	=
	\argmax_{\ket{\omega} \in \mathcal{B}_1}  \bra{\omega}\rho\ket{\omega},
\end{align}
which is determined by performing $N$ measurements for each projector. Initialize the learning eigenvalue as $q_0 =0$.

	\item[(S2)]
	Set a random vector $\pmb{\delta}_{k}$ of dimension $d$ whose entries are obtained from the set $\left\{\pm 1 \pm \sqrt{-1}\right\}$ with equal probabilities for the $k$th iteration ($k=1,2,\ldots$) and use the two gain parameters \cite{ATND:19:SR,SP:98:ITAC}: 
	\begin{align}
		\alpha_{k}&=\frac{1}{k^{0.602}} \\
		\beta_{k}&=\frac{0.1}{k^{0.101}}.
	\end{align}
	
	\item[(S3)]
	Form  two sets $\mathcal{S}_k^\pm=\left\{\ket{\eta_k^\pm}\bra{\eta_k^\pm} , I -\ket{\eta_k^\pm}\bra{\eta_k^\pm} \right\}$ of measurement operators for the $k$th iteration where
	\begin{align}
		\ket{\eta_k^\pm}
		&=
		\frac{
		\mathbf{x}_k^\pm
		}{
		\| \mathbf{x}_k ^\pm\|
		}, 
		\\
		\mathbf{x}_k^\pm
		&=
		\ket{\phi_{k-1}}
		\pm
		\beta_k 
		\pmb{\delta}_{k}.
	\end{align}

	\item[(S4)] Obtain the gradient vector in the direction of $\pmb{\delta}_{k}$ as follows:
	\begin{equation} \label{gradient}
		\mathbf{g}\left(\ket{\phi_{k-1}}\right)=\left(\frac{ \mu_k^{+}-\mu_k^{-}}{2 \beta_{k}}\right)\pmb{\delta}_{k}
	\end{equation}
	where 
	\begin{align}	\label{eq:mu}
		\mu_k^{\pm}
		=
		\bra{\eta_k^\pm}
		\rho
		\ket{\eta_k^\pm},
	\end{align}
	which is estimated by performing $N$ measurements for each projector. 
		
	\item[(S5)] Update the unit-norm learning vector and the learning eigenvalue as follows:
	\begin{align}
		\ket{\phi_{k}}
		&=
		\frac{
		\mathbf{y}_k
		}{
		\| \mathbf{y}_k \|
		}, 
		\\
		\mathbf{y}_k
		&=
		\ket{\phi_{k-1}}+\alpha_{k} \mathbf{g}\left(\ket{\phi_{k-1}}\right),
		\\
		q_k
		&=
		q_{k-1}+
		\bra{\eta_k^{+}}\rho\ket{\eta_k^{+}}+\bra{\eta_k^{-}}\rho\ket{\eta_k^{-}}.\label{eignvalue}
	\end{align}
	
	\item[(S6)] Repeat the steps (S2)--(S5) up to $k=K$. The updated learning vector produces the estimate $\ket{\hat{\psi}_{1}}=\ket{\phi_{K}}$ of the first eigenvector corresponding to the largest eigenvalue of $\rho$, which converges to $\ket{\psi_{1}}$ as $K \rightarrow \infty$.
	
	\item[(S7)] Obtain the estimate $\hat{p}_1 = q_K/\left(2K\right)$ of the largest eigenvalue using a  sample average approximation.
		
	\item[(S8)]Initialize the learning state vector  $\ket{\phi_{0}}$ to estimate the $i$th eigenvector $\ket{\psi_{i}}$, $i \geq 2$:
	We generate any random vector $\ket{v}$ and select its projection on the intersection of nullspaces of previously estimated eigenvectors, i.e,
\begin{align}
	\ket{\phi_{0}}
	=
	P_{i}\ket{v},
\end{align}
where $P_{i}=I-\sum_{\ell = 1}^{i - 1}\ket{\hat{\psi}_{\ell}}\bra{\hat{\psi}_{\ell}}$ is the projector on the nullspace of already estimated eigenstates.

\alglanguage{pseudocode}
\begin{algorithm}[t!]
	\small
	\caption{Self-guided quantum state learning for mixed states}
	\label{Algorithm:OQSL}
	\KwInput{Initializing set $\mathcal{B}_1$ of MU bases and $N_\mathrm{max}$ copies of the density matrix $\rho$}
	\KwOutput{Estimated density matrix $\hat{\rho}$}
	
	$i\leftarrow 1$ \\
	\While{$i \leq d $} {
		\eIf{$i=1$}{
			$\ket{\phi_{0}}
				\leftarrow
				\argmax_{\ket{\omega} \in \mathcal{B}_1}  \bra{\omega}\rho\ket{\omega}
				: \text{$N$ measurements for each projector}
			$ 
		}{
			$P_{i}\leftarrow I-\sum_{\ell = 1}^{i - 1}\ket{\hat{\psi}_{\ell}}\bra{\hat{\psi}_{\ell}}
			$ \\ 
			$\ket{\phi_{0}}
				\leftarrow
				P_{i}\ket{v}
			$
		}
		$q_0 \leftarrow 0$ \\		
		\For{$k \in 1 \rightarrow K$}{			
			Construct vector $\pmb{\delta}_{k}$ whose entries are chosen randomly from the set $\left\{\pm 1 \pm \sqrt{-1}\right\}$ \\
			$\alpha_{k} \leftarrow \frac{1}{k^{0.602}}$ \\
			$\beta_{k} \leftarrow \frac{0.1}{k^{0.101}}$\\		
			$\ket{\eta_k^\pm} \leftarrow$  normalized $\ket{\phi_{k-1}}\pm\beta_{k}\pmb{\delta}_{k}$ \\
			$\mu_k^{\pm}
				\leftarrow 
				\bra{\eta_k^\pm}
				\rho
				\ket{\eta_k^\pm}
				-	
				\sum_{j=1}^{i-1}
					\bigl|
					\braket{\eta_k^\pm |\hat{\psi}_{j}}
					\bigr|^{2}
				: \text{$N$ measurements for each projector}
			$ \\
			$\mathbf{g}\left(\ket{\phi_{k-1}}\right)\leftarrow\left(\frac{ \mu_k^{+}-\mu_k^{-}}{2 \beta_{k}}\right) \pmb{\delta}_{k}$ \\ 
			$\ket{\phi_{k}} \leftarrow$ normalized $\ket{\phi_{k-1}}+\alpha_{k} \mathbf{g}\left(\ket{\phi_{k-1}}\right)$ \\
			$q_k \leftarrow q_{k-1} + \bra{\eta_k^{+}}\rho\ket{\eta_k^{+}}+\bra{\eta_k^{-}}\rho\ket{\eta_k^{-}}$			}
		$\ket{\hat{\psi}_{i}} \leftarrow \ket{\phi_{K}}$ \\
		$\hat{p}_i \leftarrow \frac{1}{2K} q_K$ \\
		$\hat{r} \leftarrow i$ \\
		\eIf{$\sum_{j=1}^{i} \hat{p}_j\leq 1-\epsilon$}{
			$i  \leftarrow i +1$
		}{
			$i\leftarrow d+1$
		}
	}
	$\begin{bmatrix} \ket{\hat{\psi}_{1}^\ast} & \ket{\hat{\psi}_{2}^\ast} & \cdots & \ket{\hat{\psi}_{\hat{r}}^\ast} \end{bmatrix} 
		\leftarrow 
		\text{qr}\bigl(
			\begin{bmatrix} \ket{\hat{\psi}_{1}} & \ket{\hat{\psi}_{2}} & \cdots & \ket{\hat{\psi}_{\hat{r}}}  \end{bmatrix}
		\bigr) 
	$ \\	
	\eIf{$\hat{r} = d$}{
		\For{$i \in 1 \rightarrow \hat{r}$}{
			$\hat{p}_i^\ast \leftarrow \hat{p}_i$
		}		
	$\hat{p}_i^\ast\leftarrow\hat{p}_i^\ast/\sum_{i} \hat{p}_i^\ast$
	}{
		\For{$i \in 1 \rightarrow \hat{r}$}{
			$
				\hat{p}_i^\ast
				\leftarrow  \bra{\hat{\psi}_{i}^\ast} \rho \ket{\hat{\psi}_{i}^\ast}
				: \text{$2NK \left( d - \hat{r} \right)$ measurement in $\mathcal{B}_{\hat{\rho}}$}
			$
		}
	}
	return $\hat{\rho}
	\leftarrow
	\sum_{i=1}^{\hat{r}}
	\hat{p}_i^\ast
	\ket{\hat{\psi}_{i}^\ast}
	\bra{\hat{\psi}_{i}^\ast} $
	
\end{algorithm}
	
	\item[(S9)] 
	Refine the quantity $\mu_k^{\pm}$ in (\ref{eq:mu}) as
	\begin{equation} \label{eigenvectors_r}
		\mu_k^{\pm}
		=
		\bra{\eta_k^\pm}
		\rho
		\ket{\eta_k^\pm}
		-	
		\sum_{j=1}^{i-1}
		\left|
		\braket{\eta_k^\pm|\hat{\psi}_{j}}
		\right|^{2}
	\end{equation}
	and iterate the steps (S2)--(S7) to get $\ket{\hat{\psi}_{i}}$ and $\hat{p}_i$ for $i \geq 2$.
	
	\item[(S10)] Stop learning a new eigenvector if the estimated eigenvalues sum up near to one such that $\sum_{i=1}^{\hat{r}} \hat{p}_i \leq 1- \epsilon$ where $\hat{r}$ is the total number of nonzero estimated eigenvalues, i.e., the estimated rank.
	
	\item[(S11)] Form the orthonormal basis 
	\begin{align}
		\mathcal{B}_{\hat{\rho}}
		= 
		\left\{
		\ket{\hat{\psi}_{i}^\ast}
		\Bigm|
		i=1,2,\ldots,\hat{r}
		\right\}
	\end{align}
	using the Gram-Schmidt process with the estimated eigenvectors $\ket{\hat{\psi}_{1}}, \ket{\hat{\psi}_{2}}, \ldots, \ket{\hat{\psi}_{\hat{r}}}$. Set the estimated eigenvalues to $\hat{p}_i^\ast =\hat{p}_i$ if $\hat{r}=d$ (full rank) and normalize them. Otherwise, estimate the eigenvalues $\left\{ \hat{p}_i^*\right\}_i$ by measuring the remaining $2NK\left( d - \hat{r} \right)$ copies of the quantum state in $	\mathcal{B}_{\hat{\rho}}$.

	\item[(S12)] Construct finally the density matrix as follows:
	\begin{align}
		\hat{\rho}
		=
		\sum_{i=1}^{\hat{r}}
		\hat{p}_i^\ast
		\ket{\hat{\psi}_{i}^\ast}
		\bra{\hat{\psi}_{i}^\ast}.
	\end{align}
	
\end{enumerate}

The pseudocode of self-guided for mixed state is given in Algorithm~\ref{Algorithm:OQSL}, which utilizes $N_{\mathrm{tot}} = N \left( d \left( 2K + 1\right) + 1 \right)$ copies of the $d$-dimensional unknown state $\rho$.

\begin{figure}[t!]
	\centering
	\includegraphics[width = 0.6 \textwidth]{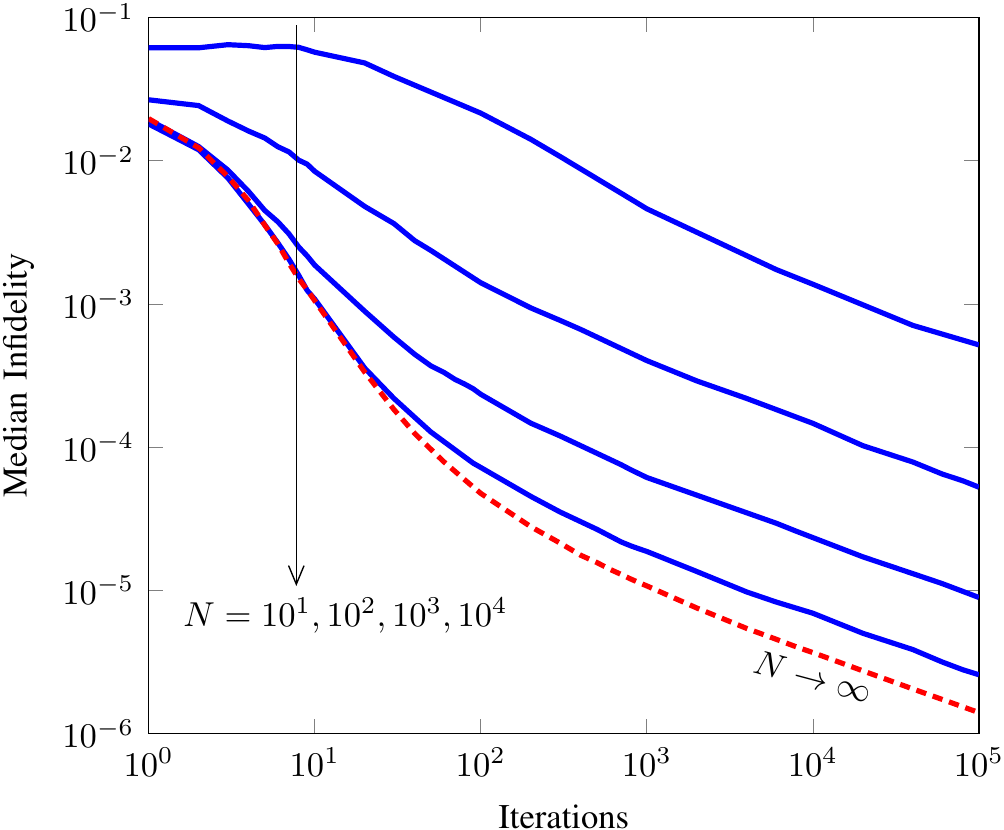}
	\caption{The median infidelity against the number of iterations achieved by our algorithm. Each blue line shows the median performance of the  $10^{3}$ randomly generated qubit states according to the Haar measure. The repetitions per measurement setting are $N=10^{1},10^{2}, 10^{3}$, and $10^{4}$.  The red dashed line shows the asymptotic performance, i.e.,  $N\rightarrow\infty$.   } \label{fig1}
\end{figure}

\section{Results}

First, we compare the computational complexity of our post-processing step with the post-processing in other well known quantum state tomography methods. The computational complexity of our algorithm is $O\left( d^3\right)$, which is due to the Gram-Schmidt decomposition \cite{AB:67:BIT}.  Since we only need to store estimated eigenvectors, the storage space requirement for our algorithm is $O\left( \hat{r} d \right)$.
	In comparison,  the standard quantum tomography has the computational complexity $O\left(12^{n}\right)$ for estimating an $d$-qubit state \cite{HZTDQLWNXLG:16:NJP,SGS:12:PRL}.  
	While SGQT offers excellent advantage in the computational complexity, i.e., no post-processing is required, it is only applicable for the tomography of pure states. 
The hybrid model PAQT is applicable on general (mixed) quantum states at the cost of increased computational complexity and storage space of the system. 
	Its computational cost depends on the Bayesian mean estimation. Bayesian mean estimation with particle filtering requires computational cost $O\left(d^4M p\right)$ for a $d\times d$ state where $M$ is the number of measurements and $p$ is the number of particles.
	Its storage space requirement is also large since it requires storing all the particles, i.e., elements in $\mathbb{C}^{d\times d}$.

To ascertain the accuracy and effectiveness of our procedure, we use infidelity as a figure of merit. The infidelity between the density matrix  $\rho$ and its estimate $\hat{\rho}$ is defined as \cite{NC:02:AA}
\begin{equation}
	\bar{F}\left(\rho,\hat{\rho}\right)
	=1-\left(\tr\sqrt{\sqrt{\rho}\hat{\rho}\sqrt{\rho}}\right)^{2} 
\end{equation} 
and its median value over the ensemble of the density matrix $\rho$ is denoted by $\bar{F}_\mathrm{med}\left(\rho,\hat{\rho}\right)$. Instead of the mean infidelity, we use the median infidelity due to its robust statistical nature that is not skewed by relatively few outliers in the estimation.
We set $\epsilon=10^{-4}$ for all examples.

Figure~\ref{fig1} shows the median infidelity $\bar{F}_\mathrm{med}\left(\rho,\hat{\rho}\right)$ as a function of the iteration number $K$ when the measurement repetitions are set to $N=10^{1},10^{2},10^{3}$ and $10^{4}$, where $10^3$ mixed qubit states $\left(d=2\right)$ are generated at random according to the Haar measure. For comparison, we also plot the asymptotic infidelity as $N\rightarrow \infty$ (red dashed line).
We can clearly see the infidelity is improving as we increase the number of iterations along with the number of measurements per iteration. 
 	
\begin{figure}[t!]
	\centering
	\includegraphics[width=0.6 \textwidth]{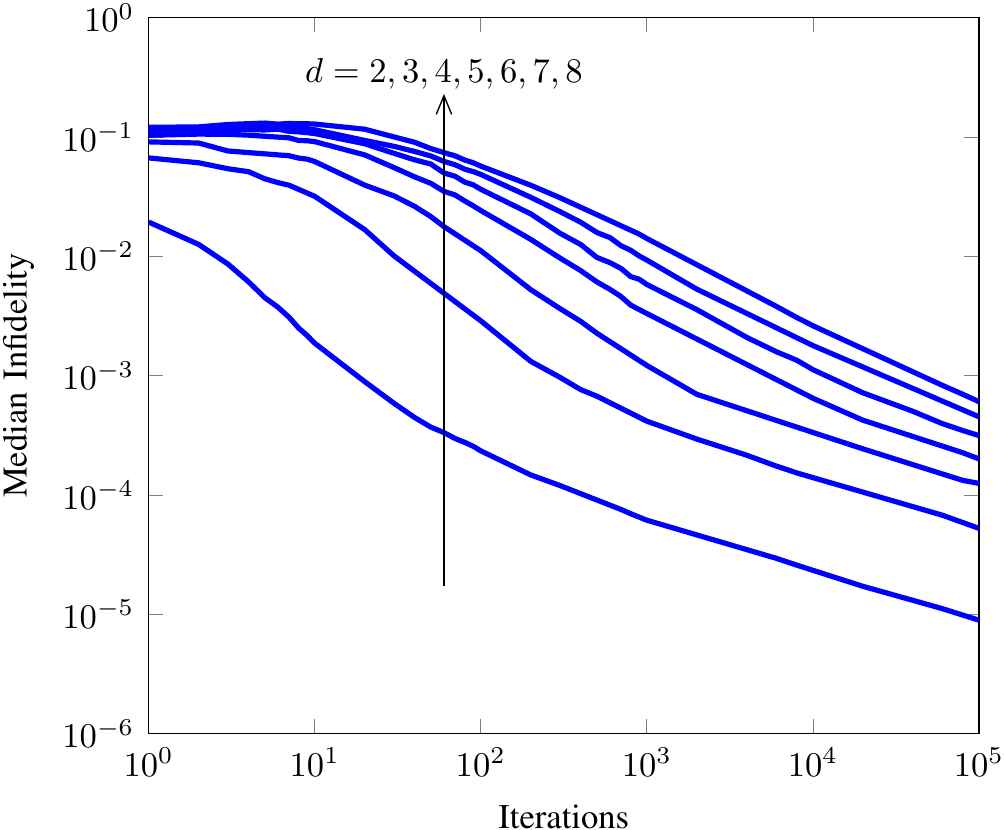}
	\caption{The median infidelity against the number of iterations achieved through our algorithm for qudit ($d \geq 2$) states. We have fixed the number of measurements per iteration $N=10^{3}$. Each line is the median performance of our procedure over $10^2$ randomly generated mixed states according to the Haar measure.  }\label{fig2}
\end{figure}

In Figure~\ref{fig2}, we plot the median infidelity of the mixed states estimated via our technique for $d=2,3,4,5,6,7,8$ against the number of iterations where we have fixed the number of measurements per iteration $N=10^{3}$.  As expected, we require more copies to achieve a given infidelity for higher dimensional mixed quantum states.

\begin{figure*}[t!]
	\centering     
	\subfigure[$d=2$]{\label{fig:a}\includegraphics[width=70mm]{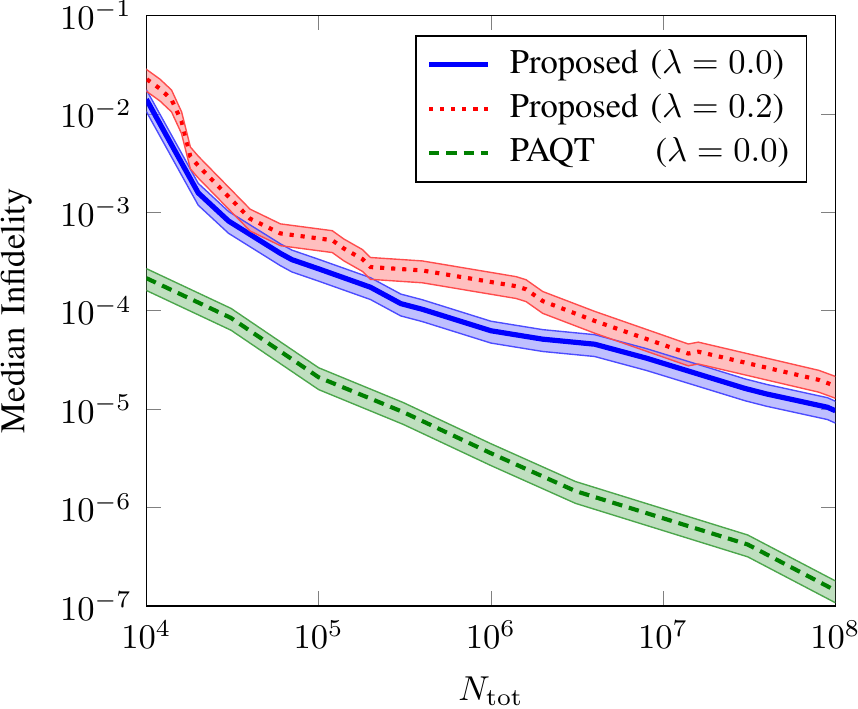}}
	\subfigure[$d=4$]{\label{fig:b}\includegraphics[width=70mm]{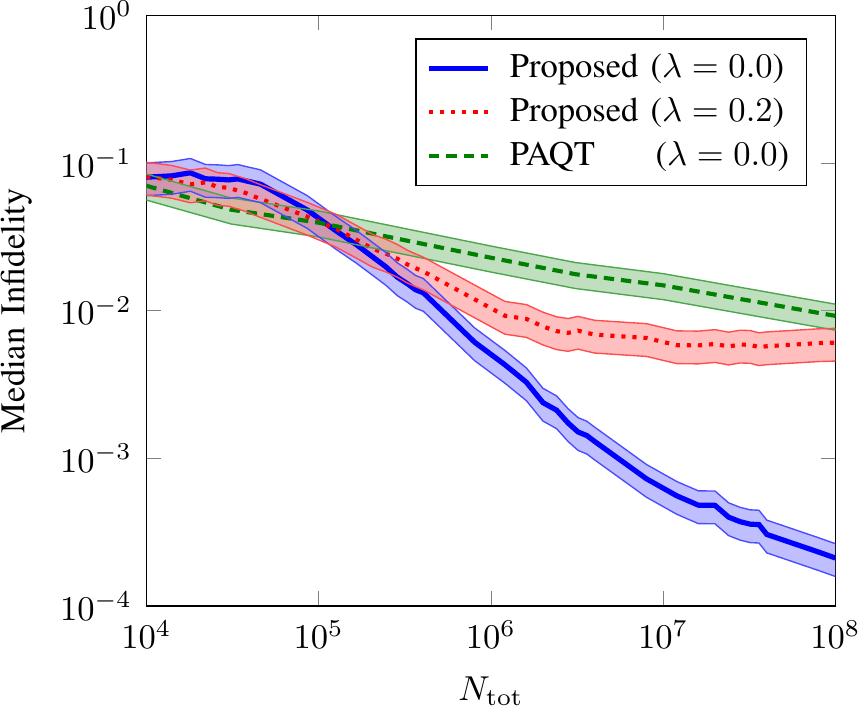}}
	\subfigure[$d=6$]{\label{fig:b}\includegraphics[width=70mm]{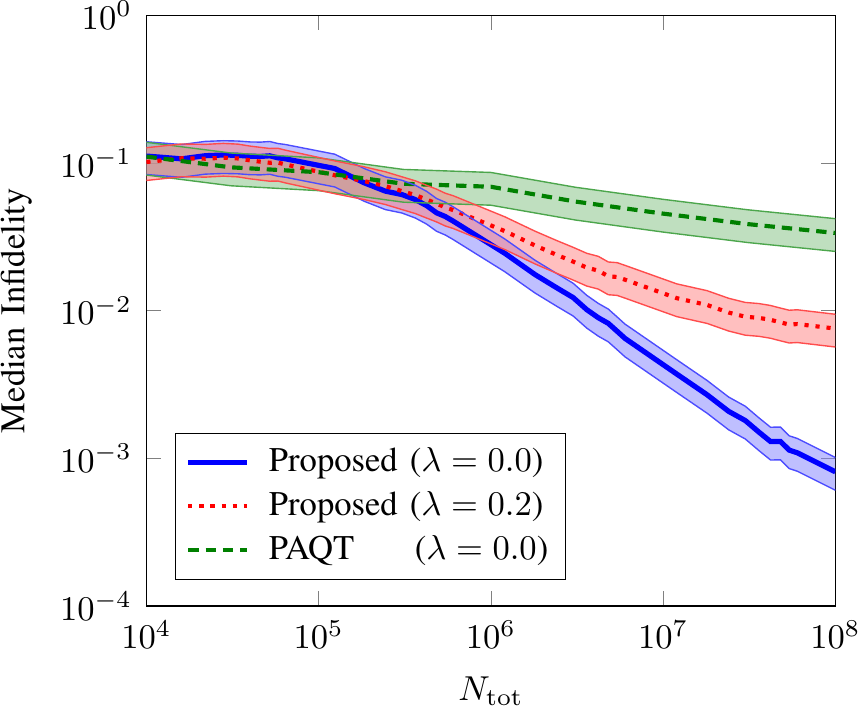}}
		\subfigure[$d=8$]{\label{fig:b}\includegraphics[width=70mm]{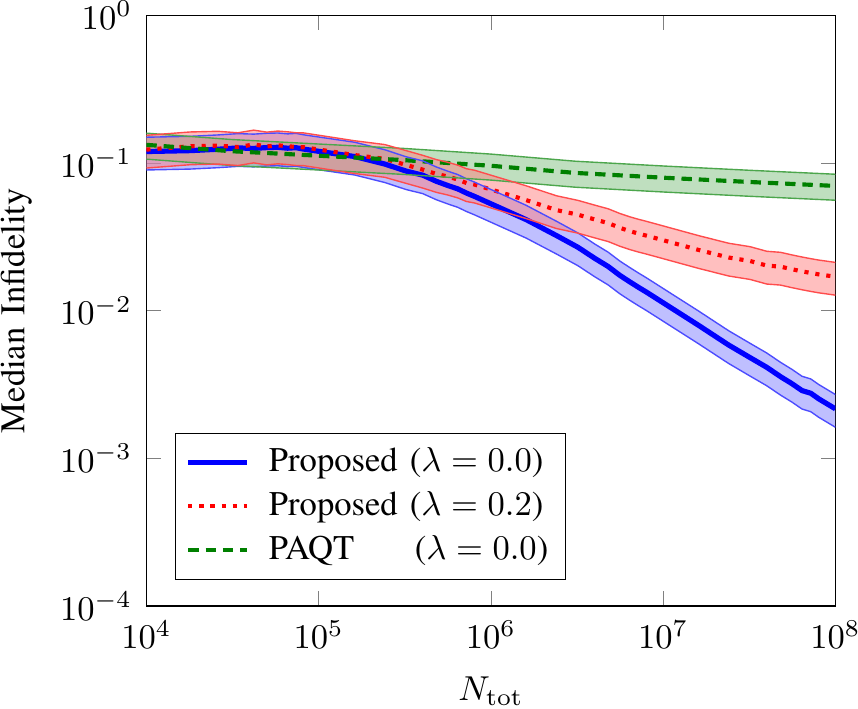}}
	\caption{The median infidelity against the number of copies achieved by proposed scheme and PAQT with measurement noise of strength $\lambda$.  Each line is the median performance over 100 randomly generated quantum states according to the Haar measure. For our algorithm, we utilize $10^4$ copies for MU bases measurement of S1. Then, we fix $N = 10^3$,  and $K = \left(N_{\mathrm{tot}} - 10^4\right)/Nd$ such that the total copies are same for all $d$. For PAQT, we use $N = 10^3$, $K = 10^5$, resampling parameter $a = 0.9$, and number of particles $p = 2000,~32000,~64000,$ and 96000 for $d = 2, 4, 6$, and 8, respectively. The shaded regions represent 25-75\% quantiles of our data. }
			\label{fig4}
\end{figure*}

Another feature of our protocol is its robustness against channel and measurement noise. To demonstrate this robustness, we perform noisy execution of our algorithm where we introduce uniform random noise in the measurement outcomes. To this end, we multiply the obtained vector of estimated probabilities with a doubly stochastic matrix $\Lambda$, which is equivalent to employing a faulty measurement device \cite{MZO:20:Quant,URS:20:QIP}. The elements of $\Lambda$ are 
$$
	\Lambda_{i, j} = \begin{cases}
		1 - \lambda + \lambda/d, &\text{when } i = j\\
		\lambda /d,  &\text{otherwise},
	\end{cases}
$$
where $\lambda$ controls the strength of introduced noise in the measurement outcomes. A noiseless measurement device corresponds to $\lambda = 0$, and a completely random measurement device has $\lambda = 1$.
This particular measurement can also be modeled as a depolarizing noise on the state before measurement, followed by a noiseless measurement device \cite{RS:arXiv:21}. 

In Figure~\ref{fig4}, we plot the ideal performance of our algorithm as well as its performance with measurement (equivalently,  depolarizing) noise with $\lambda = 0.2$. For comparison, we also plot the \emph{noiseless} performance of PAQT. 
Since the type of noise (depolarizing) we introduce in our system pushes the eigenvalues of the estimated state to $1/d$, the largest eigenvalue of the noiseless state cannot be smaller than the estimated largest eigenvalue. For this reason, we omit the first estimated eigenvalue from the normalization step. That is, let $S = \sum_i \hat{\lambda}_i$, we normalize the set of eigenvalues as $\lambda_1^* =  \hat{\lambda}_1$, and $\lambda_i^* = \hat{\lambda}_i/S$ for $1 < i \leq t$, where $t$ is the largest number such that $\sum_{i = 1}^t \lambda_i^* \leq 1$. 
There are two key insights from the analysis of Figure~\ref{fig4}. First, the relative advantage of our algorithm over PAQT becomes more pronounced as we increase the dimension $d$ of the quantum state. Second, the proposed scheme continues to outperform noiseless PAQT for higher $d$, even when a uniform noise of moderate strength $\lambda = 0.2$ is introduced in our algorithm.

\section{Experimental Results}
{We experimentally demonstrate our algorithm on IBMQ devices \cite{IBM:20:BE}. We have run our experiments on \texttt{ibmqx2} processor that has average {CNOT} and readout errors of $1.577e^{-2}$ and $4.538e^{-2}$, respectively. }
 	
{In order to prepare mixed qubit states on the IBMQ devices, we prepare a purification on two qubits of the intended state followed by the tomography on only one of the qubits. The purification of a density matrix $\rho^{A}$ that has the eigendecomposition $ \sum_{i}p_i \ket{\psi_i}\bra{\psi_i}$ is given by the composite system
 \begin{align}
 	\ket{\psi}_{AR}=\sum_{i}\sqrt{p_{i}}\ket{\psi_{i}}_{A}\otimes \ket{r_{i}}_{R},
\end{align}
where $\ket{r_{i}}_R$ is an orthonormal basis of the reference system $R$. Then,  the original mixed state is related with the composite system as
\begin{align}
\rho^{A}=\tr_{R}\left(\ket{\psi}_{AR}\bra{\psi}_{AR}\right).	
\end{align}}
\begin{figure}[t!]
	\centering
	\includegraphics[width=0.6 \textwidth]{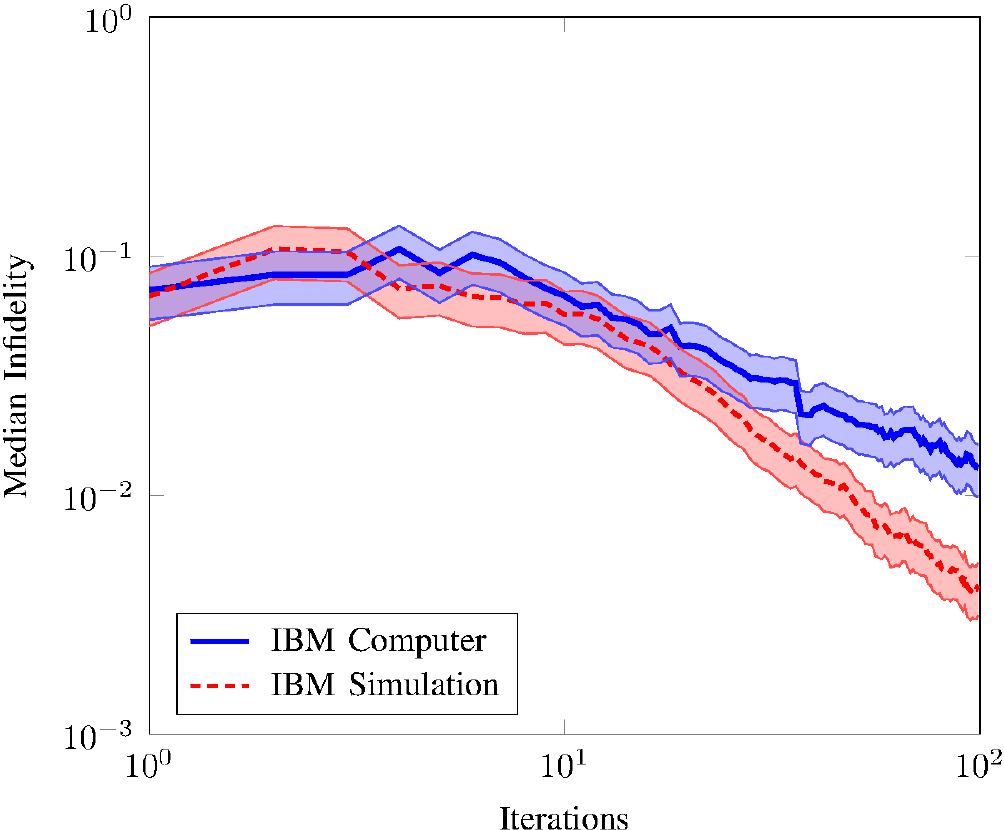}
	\caption{The median infidelity vs the number of iterations achieved by proposed algorithm on IBMQ device and Simulator. Both lines represent the median infidelity over 50 randomly generated mixed states according to the Haar measure. The shaded region represents $25\%$ quantile over the trials. We used 100 shots per iteration. }\label{fig6}
\end{figure}

For our algorithm, the bases are changing in each iteration. To measure the quantum state in desired basis, we apply unitary on the given quantum state.  The lower panel in Figure~\ref{fig:1} shows the circuit that we implemented on the IBMQ device where 
\begin{align}
H=\frac{1}{\sqrt{2}}\begin{bmatrix}
	1&1\\
	1&-1
\end{bmatrix}, \hspace{1cm}
S^{\dagger}=\begin{bmatrix}
	1&0\\
	0&e^{-\frac{\dot{\iota} \pi}{2}}
\end{bmatrix},
\end{align} 
and the unitary $U_i$ is the general unitary denoted by $u3$ in the IBMQ programming framework. 
Its expression is given by
\begin{align}
u3\left( \theta, \phi, \lambda\right) =\begin{bmatrix}
	\cos\left(\frac{\theta}{2}\right) &-e^{\dot{\iota} \lambda} \sin\left(\frac{\theta}{2}\right)\\
	e^{\dot{\iota} \phi} \sin\left(\frac{\theta}{2}\right) &e^{\dot{\iota} \left(\phi+\lambda\right)} \cos\left(\frac{\theta}{2}\right)
\end{bmatrix},
\end{align} 
where $\theta$, $\phi$ and $\lambda$ are determined by orthonormal basis in which we want to measure the state.

Figure~\ref{fig6} shows the performance of our algorithm on IBM computer and IBM simulator. IBM computer performance deviates from the simulator due to the state preparation, gate, and measurement errors in the device.

\section{Conclusion}\label{sec 4}

We have provided an online learning algorithm for quantum state tomography. Our algorithm features iterative applications of SPSA for resource efficient and accurate estimates of an unknown quantum state in higher dimensions. Numerical examples demonstrate the better performance as compared to the contemporary online learning algorithms. In particular, our proposal demonstrates a better robustness against measurement noise, which makes it suitable for near-term applications. Future works may include photonic implementations of the proposed algorithm to further analyze its practical advantages. In addition, our proposed algorithm can be utilized for efficient approximate eigendecomposition of
Hermitian matrices.

\section*{Acknowledgments}
This work was upheld by the National Research Foundation of Korea (NRF) give financed by the Korea government (MSIT) (No. 2019R1A2C2007037)
\bibliographystyle{unsrtnat}

\end{document}